\documentclass[trackchanges,twocolumn]{aastex7}
\usepackage{multirow}

\begin{document}

\title{A XRISM View of Relativistic Reflection in Cygnus X-1}

\correspondingauthor{Paul A. Draghis}
\author[0000-0002-2218-2306]{Paul A. Draghis}
\altaffiliation{MIT Kavli Postdoctoral Fellow}
\affiliation{MIT Kavli Institute for Astrophysics and Space Research, Massachusetts Institute of Technology, 70 Vassar St, Cambridge, MA 02139, USA}
\email[show]{pdraghis@mit.edu}

\author[0000-0003-2869-7682]{Jon M. Miller}
\affiliation{Department of Astronomy, University of Michigan, 1085 South University Avenue, Ann Arbor, MI 48109, USA}
\email{jonmm@umich.edu}

\author[0000-0003-0172-0854]{Erin Kara}
\affiliation{MIT Kavli Institute for Astrophysics and Space Research, Massachusetts Institute of Technology, 70 Vassar St, Cambridge, MA 02139, USA}
\affiliation{Department of Physics, Massachusetts Institute of Technology, 70 Vassar St, Cambridge, MA 02139, USA}
\email{ekara@mit.edu}

\author[0000-0001-8470-749X]{Elisa Costantini}
\affiliation{SRON Space Research Organisation Netherlands, Niels Bohrweg 4, 2333CA, Leiden, The Netherlands}
\email{e.costantini@sron.nl}

\author[0000-0002-5966-4210]{Oluwashina Adegoke}
\affiliation{Cahill Center for Astronomy \& Astrophysics, California Institute of Technology, Pasadena, CA 91125, USA}
\email{oadegoke@caltech.edu}

\author[0000-0003-3828-2448]{Javier A. Garc\'ia}
\affiliation{X-ray Astrophysics Laboratory, NASA Goddard Space Flight Center, Greenbelt, MD, USA}
\affiliation{Cahill Center for Astronomy \& Astrophysics, California Institute of Technology, Pasadena, CA 91125, USA}
\email{javier.a.garciamartinez@nasa.gov}

%\collaboration{all}{The XRISM collaboration}

\begin{abstract}
We present the first high-resolution XRISM/Resolve view of the relativistically broadened Fe K line in Cygnus X-1. The data clearly separate the relativistic broad line from the underlying continuum and from narrow emission and absorption features in the Fe band. The unprecedented spectral resolution in the Fe K band clearly demonstrates that the flux excess can be attributed to a single, broad feature, as opposed to a superposition of previously unresolved narrow features. This broad feature can be best interpreted as emission consistent with an origin near the innermost stable circular orbit around a rapidly rotating black hole. By modeling the shape of the broad line, we find a black hole spin of $a\simeq0.98$ and an inclination of the inner accretion disk of $\theta\simeq63^\circ$. The spin is consistent with prior reflection studies, reaffirming the robustness of past spin measurements using the relativistic reflection method. The measured inclination provides reinforcing evidence of a disk-orbit misalignment in Cygnus X-1. These results highlight the unique abilities of XRISM in separating overlapping spectral features and providing constraints on the geometry of accretion in X-ray binaries. 
\end{abstract}

%\keywords{\uat{High Energy astrophysics}{739} --- \uat{Interstellar medium}{847} }

\section{Introduction} \label{sec:intro}

The high-mass X-ray binary (HMXB) Cygnus X-1 has served as the ideal laboratory for testing theories regarding the physics of accretion throughout the age of X-ray astronomy.  Harboring the most massive BH known in an X-ray binary, with a mass of $21.2\pm2.2\; M_\odot$ in orbit around a $40^{+7.7}_{-7.1}\;M_\odot$ blue supergiant star (\citealt{2021Sci...371.1046M}), its persistent accretion and proximity ($\sim 2\;\rm kpc$, \citealt{2011ApJ...742...83R, 2021Sci...371.1046M}) make it one of the brightest and most studied stellar-mass black holes (BH) in the Galaxy.

Recently, \cite{2025A&A...698A..37R} measured a smaller BH and companion mass in Cygnus X-1, and argued that the companion star is nearly filling its Roche lobe, making Cygnus X-1 a possible link between low-mass X-ray binaries (LMXBs) fed through Roche lobe overflow and wind-fed HMXBs. Recent studies suggest that a small fraction of BH HMXBs would evolve into a binary black hole (BBH) or black hole - neutron star merger observable through gravitational waves (\citealt{2023ApJ...946....4L, 2023MNRAS.524..245R}), with \cite{2021ApJ...908..118N} estimating that Cygnus X-1 has a 0.1\% probability of merging in 14 Gyr. However, based on their updated measurement, \cite{2025A&A...698A..37R} argued that Cygnus X-1 is likely to undergo a BBH merger in the next 5 Gyr, linking the seemingly distinct populations of BHs seen through gravitational waves and X-rays (\citealt{2022ApJ...929L..26F, 2025ApJ...989..227D}). 

%paragraph about inclinations, NuSTAR results, IXPE work, Chandra studies on composite line, maybe radio stuff?
Chandra studies of Cygnus X-1 revealed a complex Fe K$\alpha$ profile, composed of a narrow component consistent with neutral Fe, and a broad component consistent with emission from the irradiated inner regions of an accretion disk (\citealt{2002ApJ...578..348M}). The broad component was confirmed through the observations of missions such as XMM-Newton (\citealt{2011A&A...533L...3D}), Suzaku (\citealt{2012MNRAS.424..217F}), and NuSTAR (\citealt{2014ApJ...780...78T}). All CCD and grating-resolution spectra point at a broadening of the Fe line, consistent with emission from the proximity of the innermost stable circular orbit (ISCO) of a near-maximally rotating BH. Further NuSTAR studies confirmed that the profile of the broad line remains consistent with near-ISCO emission in both the soft (\citealt{2016ApJ...826...87W}) and the hard spectral state (\citealt{2015ApJ...808....9P}) of the source. 

However, several studies argued for alternative interpretations for explaining the origin of the observed broad Fe line. For example, \cite{1999dicb.conf..251D} and \cite{2017MNRAS.472.4220B} argue that the spectra can be similarly fit with models that require the accretion disk to be truncated, not extending to the ISCO. \cite{2018NatAs...2..652C} argues that in the hard state of Cygnus X-1, X-ray polarization studies favor the presence of an extended corona and a truncated accretion disk, or a compact corona located far from the BH. Chandra HETG studies (see, e.g., \citealt{2009ApJ...690..330H, 2016A&A...590A.114M, 2019A&A...626A..64H}) demonstrate the complexity and variability of the absorption features present in Cygnus X-1, hinting at potential impacts on the ability to constrain the broad profile of the Fe line. \cite{2011ApJ...728...13N} demonstrates that the inferred inner disk radius is dependent on the assumptions of the underlying continuum, and \cite{2018ApJ...855....3T} argues that the assumed density of the accretion disk can impact the measured inner disk radius, and therefore the inferred degree of disk truncation or the measurement of BH spin. Furthermore, works such as \cite{2008PASJ...60..585M} and \cite{2017PASJ...69...36K} used Suzaku spectra of Cygnus X-1 to argue that the inclusion of additional thermal Comptonization components in the spectral models can lead to a narrower measurement of the line profile (translating to either a truncated disk or a lower BH spin) and a lower BH spin measured through the continuum fitting method. All these works demonstrate the need for spectra with high sensitivity and high energy resolution to obtain a definitive measurement of the profile of the relativistically broadened Fe line in Cygnus X-1 and to constrain the spin of the BH and the geometry of the system.

The newly launched XRISM telescope (\citealt{2020SPIE11444E..22T}) observed Cygnus X-1 during its first months of operation, under ObsID 300049010, for an effective exposure of $\sim 125 \;\rm ks$. The observation spanned nearly three days, or about half of the 5.599829-day orbital period of the system, starting around an orbital phase of $\phi_{\rm orb}=0.65$ calculated using an MJD ephemeris of 41874.207 (\citealt{1999A&A...343..861B}). The data of this observation were first analyzed in \cite{2025PASJ..tmp..103Y}, who present a thorough description of the specifics of the observation, the data reduction, and a careful analysis of the absorption features present in the spectra. 

In this work, we expand the analysis by focusing on narrow and broad emission features in the Fe region of the spectra, with the aim of characterizing the gravitational influence of the BH on the emission from the innermost regions of the accretion disk. The paper is structured as follows. In Section \ref{sec:data} we explain the spectral extraction of the XRISM Resolve observation. In Section \ref{sec:analysis} we analyze the spectra, and we summarize our findings in Section \ref{sec:results}.

\section{Data}\label{sec:data}
The observation and data reduction are presented in detail in \cite{2025PASJ..tmp..103Y}. We re-extracted the Resolve spectra from the observation using the calibration tools in Heasoft v6.34 and CALDB version 20241115. We excluded pixels 12 (calibration) and 27 from the spectra extraction, and only included High-resolution primary events (Hp). We excluded the low-resolution secondary (Ls) events when running the \texttt{rslmkrmf} command to generate small (S) and extra-large (XL) \texttt{rmf} files. For each spectrum extracted, we generated new exposure maps through the \texttt{xaexpmap} command, which were then used as inputs for generating the \texttt{arf} files using the \texttt{xaarfgen} command. Throughout the analysis, we used the S \texttt{rmf} for exploration of the parameter space, and the regions of the parameter space with the best statistics were further explored using the XL \texttt{rmf}. All parameter values and uncertainties reported here were obtained using the XL \texttt{rmf}. All spectra were rebinned using the \texttt{optimal} binning scheme (\citealt{2016A&A...587A.151K}) through the \texttt{ftgrouppha} ftool.

\begin{figure}
    \centering
    \includegraphics[width=0.95\linewidth]{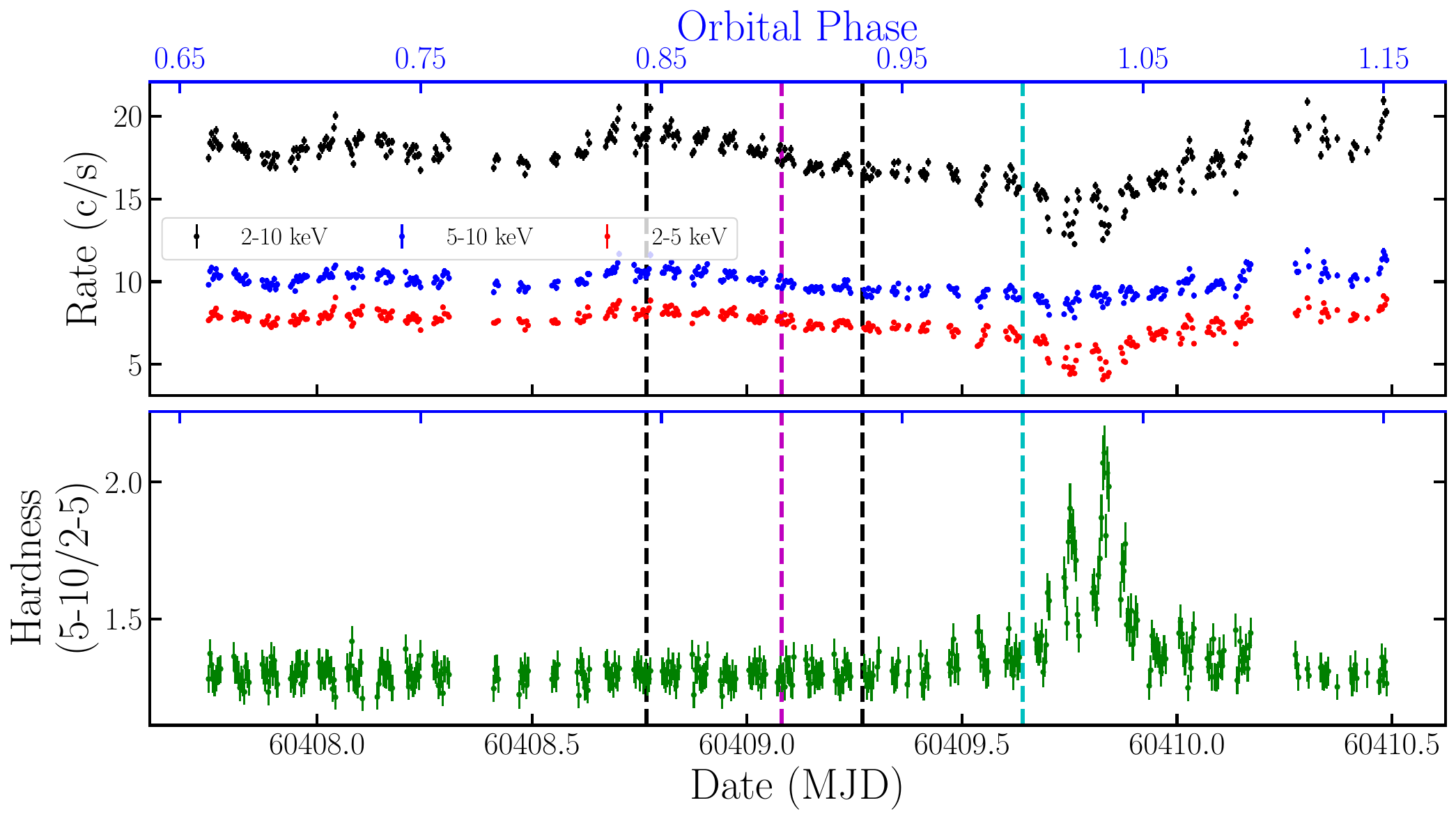}
    \caption{Top: Light curve of Cygnus X-1 as seen by XRISM Resolve during ObsID 300049010. The black, blue, and red indicate the light curves in the 2-10 keV, 5-10 keV, and 2-5 keV bands, respectively. The bottom axis shows the MJD of the observation, while the top axis shows the orbital phase of Cygnus X-1. The light curve was binned so that each point corresponds to a 300s interval. The vertical magenta and cyan lines indicate orbital phases 0.9 and 1, respectively. The vertical black lines indicate the duration of the NuSTAR observation used in this analysis. Bottom: Hardness ratio computed by dividing the 5-10 keV light curve in the top panel by the 2-5 keV light curve.}
    \label{fig:lc}
\end{figure}

The light curves of the observation in the 2-10, 2-5, and 5-10 keV bands are shown in the top panel of Figure \ref{fig:lc} in black, red, and blue, respectively, with the corresponding orbital phase computed according to the ephemeris in \cite{1999A&A...343..861B} on the top axis. As presented in \cite{2025PASJ..tmp..103Y}, the source count rate decreased during the second half of the observation. The bottom panel of Figure \ref{fig:lc} shows the hardness ratio of the source during the observation, calculated as the ratio of the 5-10 to 2-5 keV light curves shown in the top panel. The hardness ratio remains generally stable throughout the observation, with the exception of the interval between orbital phases 1.0-1.05, when the ionized winds from the massive star partially obscure the line of sight to the accretion disk leading to the onset of absorption features in the spectra. Note that this change happens after superior conjunction, when the supermassive stellar companion passes in front of the BH (phase 1, indicated by the cyan line in Figure \ref{fig:lc}).

To assess the impact of the obscuration on our ability to constrain the shape of spectral features, in order to maintain high signal-to-noise ratios (SNR), we divided the observation into two equal segments, at an orbital phase of 0.9 (magenta line in Figure \ref{fig:lc}). We extracted Resolve spectra from the two intervals, and the analysis of those spectra is presented in Section \ref{sec:analysis}. 

Furthermore, to assess the impact of including high-energy spectral coverage and the improvement over measurements made using CCD-resolution spectra, we used a 17 ks. NuSTAR (\citealt{2013ApJ...770..103H}) observation taken simultaneously with the XRISM observation (ObsID 30901039002). We extracted source and background spectra from circular regions with 180'' radius, using the standard techniques in \texttt{nustardas} v2.1.2 and the associated CALDB files in v20230530. Lastly, we rebinned the NuSTAR FPMA and FPMB spectra using the \texttt{optimal} binning scheme. The black vertical lines in Figure \ref{fig:lc} indicate the start and end times of the NuSTAR observation.

\section{Analysis}\label{sec:analysis}

\begin{figure*}[ht]
    \centering
    \includegraphics[width=0.9\linewidth]{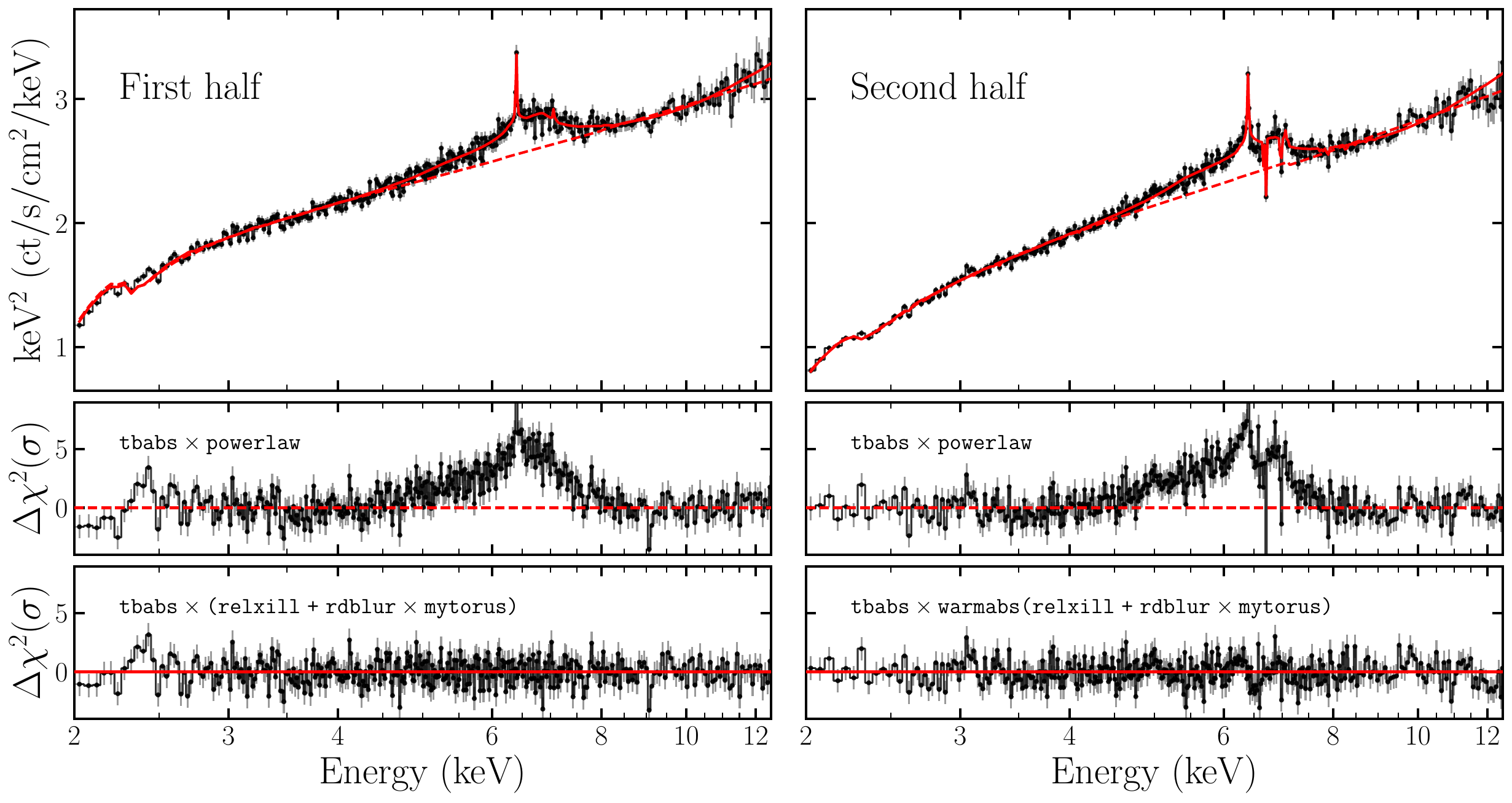}
    \caption{Spectra (top panels) and residuals produced when fitting the spectra from the first half (left) and second half (right) of the XRISM Resolve observation of Cygnus X-1 with a power-law model (middle panels) and with our best-fit model (lower panels). The solid line represents our best-fit model, while the dashed line shows the underlying power law component. The spectra in the figure were rebinned for visual clarity to have a minimum significance of 50$\sigma$ per bin, or to combine a maximum number of 20 spectral bins.}
    \label{fig:spectra_broad}
\end{figure*}

\begin{figure*}[ht]
    \centering
    \includegraphics[width=0.9\linewidth]{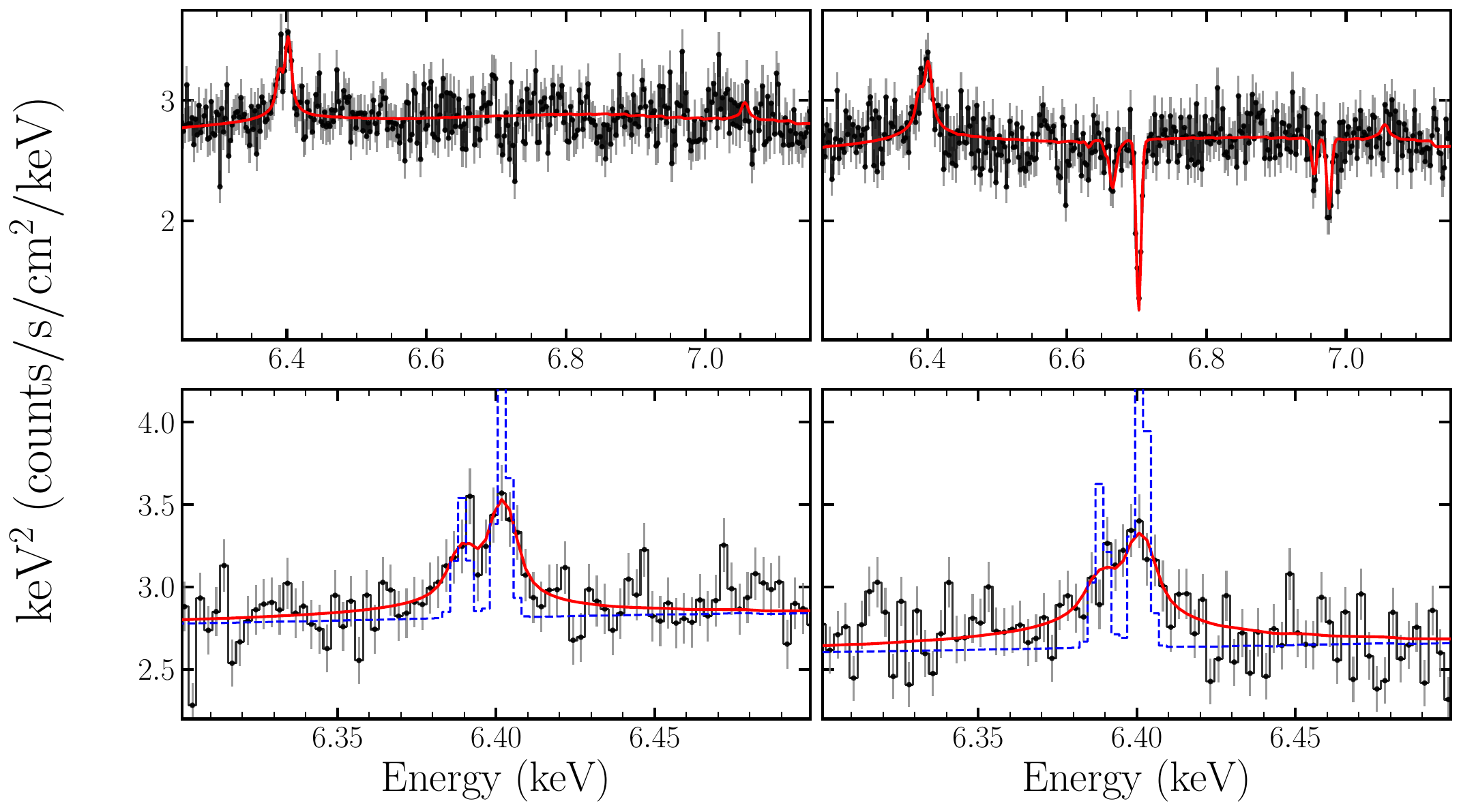}
    \caption{Focused look in the 6.25-7.15~keV (top) and 6.3-6.5 keV (bottom) at the XRISM Resolve observation of Cygnus X-1, during the first (left) and second (right) halves of the observation. The solid red lines represent our best-fit models. The dashed blue lines show the un-blurred \texttt{MYTorus} component, illustrating that the Fe K$\alpha$ complex is resolved in the data. The spectra were not rebinned for visual purposes, and display the spectral bins produced by the ``optimal" binning scheme used in the analysis.}
    \label{fig:spectra_narrow}
\end{figure*}

The top panels in Figure \ref{fig:spectra_broad} show the spectra obtained from the two halves of the observation, split below (left) and above (right) the midpoint of the observation at orbital phase 0.9. In this figure, the spectra are heavily rebinned for visual clarity. The red dashed lines represent fits to the spectra using a \texttt{powerlaw} model, and the central panels in Figure \ref{fig:spectra_broad} show the residuals in terms of $\sigma$ produced by this fit. The residuals indicate clear, irrefutable signs of the presence of a relativistically broadened Fe line. In addition to the broad line, the spectra show the presence of the seemingly stable narrow Fe K$\alpha$ line at 6.4 keV, and the emergence of absorption features in the second half of the observation. Figure \ref{fig:spectra_narrow} focuses on the region around these narrow features. The top panels show the optimally binned spectra in the 6.25-7.15 keV band and highlight the emergence of the absorption features in the data. Furthermore, the Fe K$\alpha$ line at 6.4 keV is noticeable in the data. A closer look at the Fe K$\alpha$ line (bottom panels in Figure \ref{fig:spectra_narrow}) shows signs of asymmetrical broadening consistent with distant reflection and perhaps shows signs of time evolution.

For the analysis in this section, we fit the XRISM Resolve spectra in the 2-12.5~keV band. The complete XRISM Resolve spectrum of Cygnus X-1 can be modeled by four components. First, the coronal emission produces a power law-like component, characterizing the broadband shape of the spectrum. Second, coronal radiation is reprocessed in the innermost regions of the accretion disk (i.e. ``reflected"), producing the broad excess in the Fe region. This is referred to as the relativistically broadened Fe line. These first two components are modeled using the \texttt{relxill} component (\citealt{2016MNRAS.462..751G}) by setting the value of the reflection fraction parameter to positive values, to ensure that the component models both the reflected spectrum and the underlying power-law coronal emission. In our fits, we used \texttt{relxill} v2.3. Models used to fit spectra of accreting X-ray binaries often include a model component to describe the thermal emission from the accretion disk. At the time of the observation, Cygnus X-1 was in a hard spectral state. Past studies have shown that the temperature of the disk during Cygnus X-1's hard state is low, $\sim0.2\;\rm keV$ (see, e.g., \citealt{2008PASJ...60..585M}). We tested the impact of including a \texttt{diskbb} component to our model and found that the quality of the fit is unchanged and that the component does not contribute significantly to the total model. Therefore, we omitted this component from the rest of the study. We account for galactic absorption along the line of sight through the \texttt{TBabs} model.

 %talk about warmabs
Third, the absorption features are modeled through the \texttt{warmabs} model (\citealt{2001ApJS..133..221K}). Given that the absorption features were analyzed in detail in \cite{2025PASJ..tmp..103Y}, we adopt a simplistic approach in our analysis. We included a single layer of warm ionized absorption (\texttt{warmabs}) in our model. We allowed the column density of the gas, the ionization parameter, the turbulent velocity, and the redshift to vary freely in our initial fits, and obtained the parameters that produce the best fit to the spectra. For the rest of the analysis, as it is unlikely that the narrow absorption features will be correlated with the emission features, we fixed the \texttt{warmabs} parameters at their best-fit values. For the first half of the observation, the \texttt{warmabs} only improves the fit statistic at the $2\sigma$ level. Therefore, for the rest of the analysis, we exclude the \texttt{warmabs} component from the fit to the first half of the observation, both because it is not statistically significant and also because it is computationally expensive to include it, even when its parameters are fixed. For the fit to the second half of the observation, we find a column density of $\sim7\times10^{21}\;\rm cm^{-2}$, an ionization parameter $\log(\xi)\sim3.3$, turbulent velocity of $\sim150\;\rm km/s$, and a blueshift of $z\sim3.94\times10^{-4}$ corresponding to a velocity shift of $\sim120\;\rm km/s$. Given that the \texttt{warmabs} model assumes a power-law ionizing spectrum of index $\Gamma=2$ and constant ionization throughout the optically thin medium, how the calculation does not include the UV ionizing flux from the companion star, and especially since the absorption spectrum was characterized in detail in a different paper, we simply limit our investigation to finding the values that reduce the fit statistic and only report these values here for completeness, without calculating their uncertainties, as these values fail to capture the entire physical picture. While the bottom-right panel of Figure \ref{fig:spectra_broad} appears to indicate significant absorption residuals in the Fe band around 6.5 keV, inspecting the spectra and models in Figure \ref{fig:spectra_narrow} indicates that the \texttt{warmabs} component produces good fits to the shapes of the Fe XXV and Fe XXVI lines.

Lastly, to model the narrow emission Fe K$\alpha$ and Fe K$\beta$ lines, we used the \texttt{MYTorus} component (\citealt{2009MNRAS.397.1549M}). We used only the line configuration of MYTorus, with attempts to utilize the complete configuration not producing statistically significant improvements in fit quality or any significant changes in fit parameters. The blue dashed lines in the bottom panels in Figure \ref{fig:spectra_narrow} show the neutral profile of the Fe K$\alpha$ doublet as predicted by \texttt{MYTorus}, illustrating that the two components are resolved by the data, and that the spectra require an additional broadening component. We first tested a Gaussian smoothing function \texttt{gsmooth}, obtaining a good fit. We further tested the effect of replacing the \texttt{gsmooth} component with a more physically motivated model through the \texttt{rdblur} component. Fitting the spectra in the narrow, 6.3-6.5 keV band, we find that \texttt{rdblur} is preferred over \texttt{gsmooth} at $3\sigma$ level in the first-half spectrum and at $4\sigma$ level in the second-half spectrum. Preliminary attempts to fit the narrow line profile with a model such as \texttt{photemis} that describes the emission from recombination and collisional excitation in a photoionized plasma or with multiple layers of reflection originating at different radii fail to produce significant statistical improvements in the fits. Given that the adopted model produces a good fit to the line profile, we continue our analysis in this work using this treatment. In characterizing the profile of the narrow, neutral emission Fe line around 6.4 keV, the \texttt{rdblur*MYTorus} component in the fit captures the subtle differences noticeable in the bottom panels of Figure \ref{fig:spectra_narrow}. Most noticeably, the fits measure consistent redshifts of $z\sim3\times10^{-4}$, corresponding to a velocity of $\sim90\;\rm km/s$. We reserve a more detailed analysis of the morphology of the narrow Fe K$\alpha$ line for a future publication. 

Our complete model, in \texttt{Xspec} parlance, is \texttt{TBabs*(relxil+rdblur*MYTorus)} for the first spectrum, and \texttt{TBabs*warmabs(relxil+rdblur*MYTorus)} for the second half of the observation (left and right panels of Figure \ref{fig:spectra_broad}). The residuals produced by these fits are shown in the bottom panels of Figure \ref{fig:spectra_broad}, and the complete model is shown in relation to the data through the solid red line in the top panel of Figure \ref{fig:spectra_broad}. Focusing on the narrow features, the top panels in Figure \ref{fig:spectra_narrow} show the spectra (black) and models (red) in the 6.25-7.15 keV bands, illustrating that the model describes the broad and narrow features well. Focusing even further in the 6.3-6.5 keV band, the bottom panels of Figure \ref{fig:spectra_narrow} show the fit to the narrow Fe K$\alpha$ emission line during the two halves of the observation. Both spectra seem to favor asymmetric broadening of the narrow line with a slightly extended red wing of the line, which is accounted for in our model through the convolution with the \texttt{rdblur} component. Furthermore, the line profile appears to change slightly, transitioning from a double-peaked structure during the first half of the observation to a single-peak structure during the second half.

%talk about relxill
All parameters in the \texttt{relxill} component were allowed to vary freely, with the exception of the redshift of the component, which was fixed at zero, the inner radius of the accretion disk, which was fixed to the size of the ISCO, and the outer radius of the accretion disk which was fixed at the maximum value allowed by the model, $1000\;\rm r_g$. The reflection fraction parameter was constrained to take only positive values to ensure that the model simultaneously accounts for relativistic reflection and for the underlying coronal emission.  As the sensitivity of the narrow line data is influenced by the high underlying continuum, spectral fits are unable to singularly constrain the geometry of the \texttt{rdblur} component convolved to \texttt{MYTorus}. Therefore, we fixed the inner radius of the \texttt{rdblur} component to the outer radius of the \texttt{relxill} component (1000 $\rm r_g$), and we fixed the outer radius of the \texttt{rdblur} component to its maximum value within the model of $10^7\;\rm r_g$. In this way, the emissivity profile of the \texttt{rdblur} component is completely determined by the emissivity index $q_M$. To further simplify the model, we linked the inclination of the \texttt{rdblur} and \texttt{MYTorus} components to that of the \texttt{relxill} component, and we linked the photon index $\Gamma$ in \texttt{MYTorus} to the equivalent parameter in \texttt{relxill}. All other parameters of the model were allowed to vary freely, including the \texttt{TBabs} column density, and all parameters of \texttt{rdblur} and \texttt{MYTorus}.

Given the complexity of the model and the data, performing simple \texttt{Xspec} error scans presents a challenging task from a computational point of view. Therefore, in order to characterize the uncertainties of the best-fit parameters, we ran MCMC simulations for the combinations of models and spectra. For each run, we allowed 120 walkers to each take 3,000 steps and discarded the first 500 steps of each walker as a burn-in phase. After marginalizing over the posterior probability distributions, we computed the modes and $\pm1\sigma$ credible intervals for each free parameter in the spectral fits, which we report in Table \ref{tab:MCMCs}. The complete corner plot obtained from the MCMC analysis is presented in Appendix \ref{sec:corner}.

\begin{deluxetable}{clcc}[!h]
%\tabletypesize{\scriptsize}
\tablecaption{MCMC results\label{tab:MCMCs}}
\tablehead{
\colhead{Comp.} & \colhead{Parameter} & \colhead{First half} & \colhead{Second half}}
\startdata
\multirow{1}{*}{\texttt{TBabs}} & $N_H\;[\times 10^{22}\rm cm^{-2}]$ & ${0.65}_{-0.02}^{+0.04}$ & ${1.7}\pm0.2$ \\\hline
\multirow{11}{*}{\texttt{relxill}} & $q_1$ & $\geq9.3$ & $\geq9.5$ \\
& $q_2$ & ${1.90}_{-0.04}^{+0.06}$ & ${1.8}_{-0.2}^{+0.1}$ \\
& $R_{\rm br}\;[\rm r_g]$ & ${3.2}\pm0.1$ & ${3.8}_{-0.5}^{+0.3}$ \\
& $a$ & ${0.975}_{-0.005}^{+0.004}$ & ${0.981}_{-0.005}^{+0.004}$ \\
& $\theta\;[^\circ]$ & ${62}\pm1$ & ${65}\pm1$ \\
& $\Gamma$ & ${1.79}\pm0.1$ & ${1.73}_{-0.02}^{+0.04}$ \\
& $\log(\xi)\;[\rm erg\;cm\;s^{-1}]$ & ${2.86}_{-0.04}^{+0.01}$ & ${2.85}_{-0.07}^{+0.06}$ \\
& $A_{\rm Fe}\;[A_\odot]$ & ${0.63}_{-0.02}^{+0.01}$ & ${0.7}_{-0.1}^{+0.2}$ \\
& $E_{\rm cut}\;[\rm keV]$ & ${260}_{-40}^{+110}$ & ${230}_{-40}^{+90}$  \\
& \texttt{refl\_frac} & ${1.7}\pm0.1$ & ${1.8}_{-0.1}^{+0.8}$ \\
& $\rm norm_{\rm R}\;[\times 10^{-3}]$ & ${26.6}_{-0.5}^{+1.4}$ & ${27}_{-6}^{+1}$ \\\hline
\multirow{1}{*}{\texttt{rdblur}} & $q_{M}$ & ${1.99}_{-0.03}^{+0.04}$ & ${2.16}_{-0.09}^{+0.05}$ \\
\multirow{1}{*}{\texttt{*}}& $N_{H,M} \;[\times 10^{22}\rm cm^{-2}]$ & ${12}_{-3}^{+10}$ & ${13}_{-3}^{+8}$ \\
\multirow{2}{*}{\texttt{MYTorus}} & $z \;[\times 10^{-5}]$ & ${23}_{-3}^{+9}$ & ${44}_{-25}^{+19}$ \\
& $\rm norm_{\rm M}$ & ${0.46}_{-0.12}^{+0.04}$ & ${0.47}_{-0.08}^{+0.17}$ \\\hline
& \multirow{2}{*}{C-Stat$/\nu$}   & ${3900}\pm6$ & ${3684}_{-4}^{+7}$         \\
&                                 & $(3888/3706)$          & $(3673/3684)$               \\ 
\enddata
\tablecomments{We report the mode and $\pm1\sigma$ credible intervals of the posterior distributions resulting from the MCMC runs. For the C-Statistic, we report the values computed based on the MCMC runs and the values obtained from direct \texttt{Xspec} fitting, along with the number of degrees of freedom, in parentheses.}
\end{deluxetable}

The fits to the two spectra measure an increased equivalent hydrogen column density through the \texttt{TBabs} component in the second spectrum. While the \texttt{TBabs} component is intended to account for galactic obscuration along the line of sight, the increase in measured column density is likely attributed to an increase in intrinsic obscuration during the second half of the observation, rather than a change in galactic structure. We attempted to incorporate prior information from the literature on the column density along the line of sight, along with measurements from simultaneous X-ray spectra taken with Xtend and NICER. However, Xtend spectra suffer from pile-up, and the Resolve spectrum shows a deviation from agreement with NICER spectra under 3 keV. Therefore, we allowed the column density in the fits to vary as a free parameter, in an attempt not to artificially bias the parameter inference.

The \texttt{relxill} parameters that most strongly shape the profile of the relativistically broadened Fe line (emissivity parameters, BH spin, viewing inclination, and ionization parameter) are consistent between the spectra from the two halves of the observation. The inner emissivity parameters take nearly maximal values over the narrow disk region between the ISCO ($\sim1.6 \; \rm r_g$ for the measured $a\simeq0.98$) and a breaking radius of $\sim3.5\;\rm r_g$, followed by a shallower emissivity profile at larger radii, with $q_2\sim1.9$. The similarity between the solutions for the two spectra strongly suggests that the fits independently converged to the same region of the parameter space, strengthening the robustness of the result. It is particularly important to note that the two spectra measure formally inconsistent inclinations at $1\sigma$. However, given that the reported uncertainties only represent the statistical uncertainty resulting from the MCMC analysis, and considering the complexity of the model and of the parameter space, it is likely that properly quantifying the systematic uncertainties associated with the measurements would formally bring the two values in agreement. In particular, because the inclination of the \texttt{relxill} component is tied to that of the \texttt{rdblur} and \texttt{MYTorus} components, it is particularly challenging to identify a singular value that simultaneously provides the best statistical fit for both the narrow and broad line. Therefore, we report that the inclination measured by this analysis is $\theta=63^\circ\pm3^\circ$, as to encapsulate both individual solutions. Similarly, we report a BH spin of $a=0.98\pm0.01$.

The spin measurement is consistent with previous reflection measurements performed on NuSTAR data (\citealt{2024ApJ...969...40D}), but the measured inclination $\theta\sim63^\circ$ is marginally higher than the $47^{+9}_{-11}$ degrees measured by \cite{2024ApJ...969...40D}. Fixing the inclination angle in the spectral fit at $47^\circ$ produces a fit worse by $\Delta C=8$ for the first half of the observation, corresponding to a difference of $\sim2.8\;\sigma$ for a single free parameter. This inclination difference can probably be attributed to the difficulty of CCD spectra with modest energy resolution to properly constrain the shape of the blue wing of the Fe line, especially when the spectra show evidence of narrow emission and absorption features in the Fe band. Alternatively, the inconsistent inclination measurement could be attributed to correlations between parameters that simultaneously influence the shape of the blue wing of the Fe line (e.g., inclination, ionization, Fe abundance; see \citealt{2025ApJ...989..227D}) and the inability to break these degeneracies due to the lack of high energy coverage that constrains the shape of the Compton hump. However, it is possible that allowing the inclination of the narrow line to differ from that of the relativistic reflection would allow for additional parameter freedom in the spectral fits. 

The remaining reflection parameters are measured to have values different between the two halves of the XRISM observation. However, this is not surprising. First, the increase in the power law index $\Gamma$ is likely to be caused by either an intrinsic change in the source spectrum or by the increase in low energy absorption during the second half of the observation. This change in $\Gamma$ is visually apparent in the upper panels of Figure \ref{fig:spectra_broad} and is corroborated by the increase in hardness shown in the lower panel of Figure \ref{fig:lc}. Second, the high-energy cutoff is inconsistent between the two spectra. Seeing how in our XRISM band pass we only included spectral bins up to an energy of 12.5~keV, it is unlikely that the parameter can be reliably constrained. Similarly, the measured Fe abundance, reflection fraction, and normalization of the component change significantly between the two spectra. Given the lack of spectral coverage of the Compton hump, it is unlikely that the degeneracy between the parameters can be broken. Still, even if the individual parameter values are possibly not accurately constrained, this combination of parameter values, through the model degeneracy between parameters, serves to produce a good spectral fit.

We attempted simultaneous fits to the spectra from the two halves of the XRISM Resolve observation, linking the parameters that are expected to remain constant in time: the equivalent hydrogen column density in \texttt{TBabs}, the BH spin, the inclination, and the Fe abundance. We maintained the same prescription of narrow emission and absorption features as throughout the paper and allowed all other parameters of the reflection spectrum to vary freely. We find that by linking parameters of interest between the models used to fit the two spectra, the model will converge to one of the values measured in individual fits and adjust other parameters to compensate for the change. In particular, the joint fit favors a column density of $\sim1.5\times10^{22}\;\rm cm^{-2}$, but adjusts the photon index $\Gamma$ and the high-energy cutoff $E_{\rm cut}$ in the model for the first half to compensate for the change at low energies. Similarly, the Fe abundance will converge to a low value of $\sim0.5$ times solar and decreases the reflection fraction and $E_{\rm cut}$ and the ionization, which in turn slightly increase the inclination to $\sim 66^\circ$. Most importantly, throughout this experiment, the measured BH spin remains similarly high, $a\simeq0.98$. It is likely that any effort to perform joint fits of the two spectra would provide some information about the comparison of the statistical and systematic uncertainties associated with the measurements. In the absence of spectral information regarding the Compton hump, the spin is most strongly determined by the shape of the relativistically broadened Fe line. However, as this experiment demonstrates, while the parameters that determine the shape of the blue wing of the Fe line can work together in a complex region of the parameter space to produce similar solutions, the red wing of the line is always similarly constrained. The blue wing is primarily determined by gas physics, whereas the red wing is primarily determined by geometry and gravity. This conclusion is supported by the study of parameter degeneracies in reflection models presented in \cite{2025ApJ...989..227D}.

To further explore the impact of the energy coverage of the Compton hump, we performed joint fits between the spectrum from the first half of the XRISM observation and a simultaneous NuSTAR observation of the source. We fit the NuSTAR FPMA/B spectra in the 8-79 keV band to ensure that the Fe band is characterized by the Resolve spectra, that the Compton hump is covered by the NuSTAR spectra, and to obtain simultaneous coverage in the 8-12.5 keV band to verify the cross-calibration of the two instruments. We maintained the same best-fit model for the spectra and included a multiplicative constant to account for calibration uncertainties between instruments. We find that the constant, which was fixed to unity for the Resolve spectrum, takes a value of $1.022$ for FPMA and $1.018$ for FPMB. Including the Compton hump helps constrain the parameters of \texttt{relxill}. In particular, the photon index decreases from $\Gamma=1.79\pm0.02$ when fitting the Resolve spectrum alone to $1.67\pm0.01$ when including NuSTAR data, the reflection fraction decreases to $0.66\pm0.05$, the Fe abundance increases to $3.9\pm0.2$, the high-energy cutoff increases to $\gtrsim800$, and the ionization increases to $\log\xi=3.01\pm0.02$. The large change in parameters suggests that the fit identified an alternative solution. However, despite changing parameters, the inferred spin and inclination remain consistent with the fit to the Resolve spectrum alone: $a=0.96\pm0.01$ and $\theta=63^\circ\pm2^\circ$.

\begin{figure}
    \centering
    \includegraphics[width=0.95\linewidth]{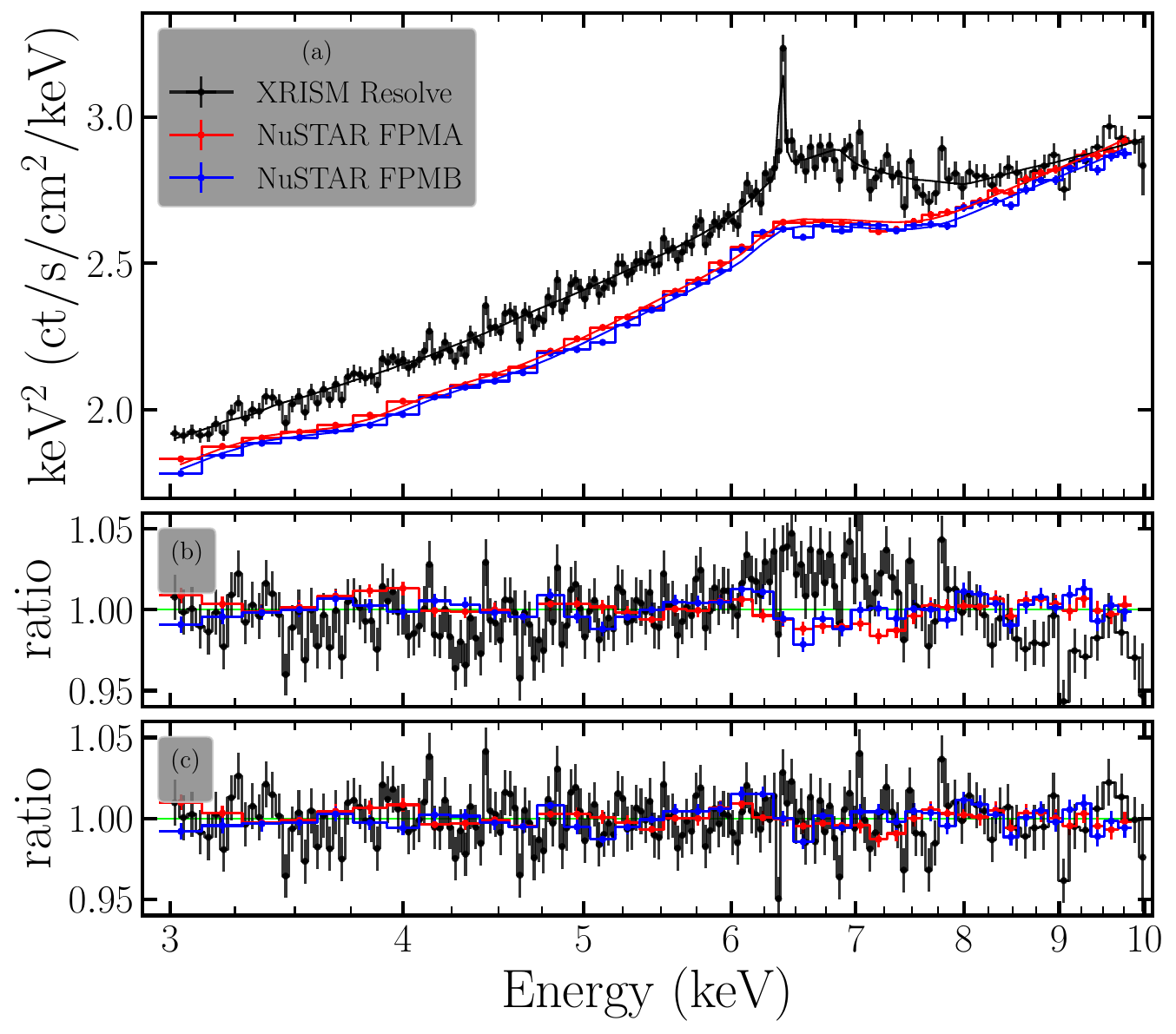}
    \caption{The top panel shows the XRISM Resolve spectrum from the first half of the observation (black) compared with simultaneous NuSTAR FPMA (red) and FPMB (blue) spectra. The bottom panels show ratio of data/model for the XRISM and NuSTAR spectra when all parameters in the spectral model are linked (panel b), and when the parameters of the reflection component are allowed to vary independently (panel c). All parameters produce agreement within $1\sigma$ between the spectra from the two instruments.}
    \label{fig:XRISM_vs_NuSTAR}
\end{figure}

As a test of the agreement between XRISM and NuSTAR spectra and of the improvement brought forth by the high-resolution of Resolve, we performed joint fits in the 3-10 keV band of spectra from both instruments with a simplified model: \texttt{constant * (relxill + gauss)}, where the \texttt{gauss} component was used to describe the narrow Fe K$\alpha$ line. These spectra are shown in the top panel of Figure \ref{fig:XRISM_vs_NuSTAR}. Given that we performed this comparison using the spectrum from the first half of the XRISM observation, we did not need to account for absorption features. This model produces a good fit for the spectra from both instruments, with a slight disagreement between the residuals produced by the two instruments (middle panel in Figure \ref{fig:XRISM_vs_NuSTAR}). To test the ability to infer the model parameters, we allowed all parameters of \texttt{relxill} to vary between XRISM and NuSTAR with the exception of the model normalization, with any flux offset being accounted for by the constant applied to the model. Additionally, we fixed the parameters of the \texttt{gaussian} component to those measured by XRISM, and tied the parameters for NuSTAR FPMA and FPMB. The residuals produced by this fit are shown in the bottom panel of Figure \ref{fig:XRISM_vs_NuSTAR}. First, we find that the overall flux offset between XRISM and NuSTAR is on the order of 2\%. Second, we find that all of the parameters in relxill agree between the two spectra within $1\sigma$. Most interestingly, the spin as measured by XRISM takes a value of $0.977\pm0.015$, while the value measured by NuSTAR is $0.976\pm0.025$. It is possible that the high-resolution of Resolve can decrease the statistical uncertainty of the measurement, by being more sensitive to small variations in spin. The largest discrepancy between parameters was found for the inclination, with XRISM measuring $\theta=59^\circ\pm3^\circ$, while NuSTAR measures $51^\circ\pm5^\circ$. It is unclear what the main driver of the different measured inclinations is, but a possible candidate is the inability of this simplistic model to capture the full extent of the way the flux of the narrow line is distributed across spectral bins when unresolved in NuSTAR data.

While the previous experiment probes the influence of high-resolution spectra on the statistical uncertainties of the spin measurement, it is possible that the systematic uncertainty associated with not accounting for unresolved narrow features in NuSTAR data has a larger magnitude. In an attempt to quantify the impact of overlooking narrow features on the ability to measure BH spin in prior-generation spectra, we performed a similar analysis on the spectrum from the entire NuSTAR bandpass of the simultaneous observation. Fitting the NuSTAR FPMA\&FPMB spectra in the 3-79 keV band with just \texttt{TBabs*relxill} produces a good fit, with no indication of residuals resembling narrow features. In this fit, the BH spin is measured to be $a=0.7\pm0.3$ and the inclination $\theta=47^\circ\pm13^\circ$. Subsequently, we added the description of the narrow Fe K$\alpha$ emission line as measured by XRISM to the model. The resulting spectral fit measures $a=0.99\pm0.01$ and $\theta=68^\circ\pm5^\circ$. Initially, when not including the narrow Fe K$\alpha$ line, the measured inclination was in good agreement with past reflection studies (see, e.g., \citealt{2024ApJ...969...40D}), and the measured spin was poorly constrained. When including the narrow line in the model, the fit produces measurements with increased precision and in good agreement with those based on Resolve data. It is important to note that this is a single experiment, using a moderately short NuSTAR observation of a single source. However, this result perhaps hints at the importance of accounting for narrow spectral features even when formally unresolved by low-resolution spectra in an attempt to characterize the systematic uncertainties associated with BH spin and inclination measurements obtained through the relativistic reflection method. This aspect is particularly relevant as an emerging number of BH XBs show evidence of narrow spectral features even in NuSTAR-resolution data, such as GX 339-4 (\citealt{2019ApJ...885...48G}) and MAXI J1820+070 (\citealt{2019Natur.565..198K, 2019MNRAS.490.1350B}).

\begin{figure*}
    \centering
    \includegraphics[width=0.9\linewidth]{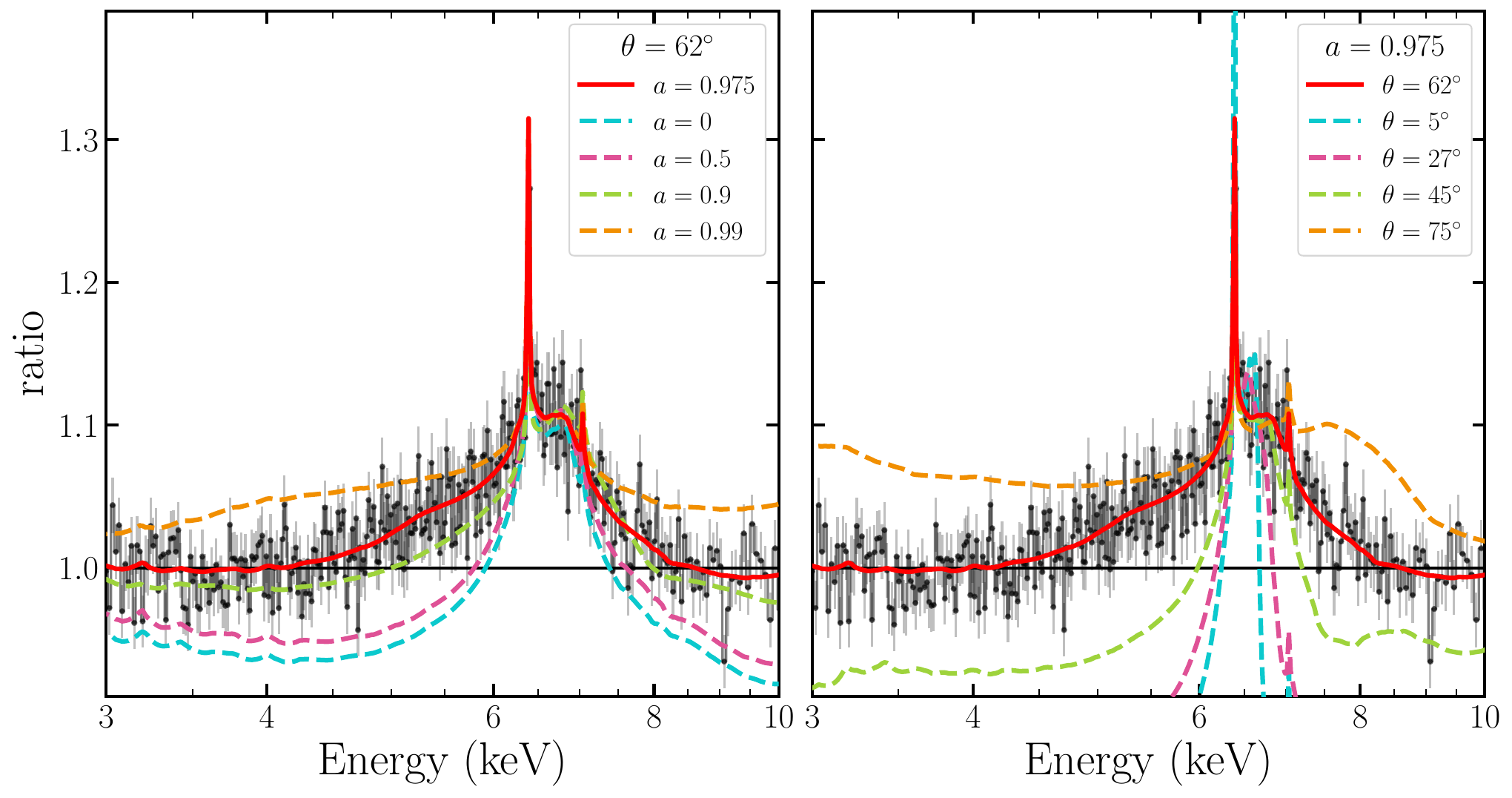}
    \caption{Ratio of spectra from the first half of the XRISM observation of Cygnus X-1 to a power law model (black line), showing the shape of the relativistically broadened Fe line. The red line in both panels shows the best-fit model, obtained with a BH spin of $a=0.975$ and an inclination of $\theta=62^\circ$. The different colored dashed lines represent the models with different BH spins (left) and different viewing inclinations (right), when the other parameter is held at the best-fit value. The models were shifted to match the flux at the peak of the line, around 7 keV, and do not represent the results of spectral fits. Lower BH spins under-predict the red wing of the Fe line in the left panel, whereas higher spins over-predict both the blue and red wings of the line. Similarly, different inclinations (right panel) fail to capture the profile of the line.}
    \label{fig:line_shape}
\end{figure*}

Figure \ref{fig:line_shape} shows the ratio of the spectrum from the first half of the observation to a simple power-law model (horizontal black line), highlighting the shape of the relativistically broadened Fe line. The solid red line represents our best-fit model for the spectrum. A few points emerge from visually inspecting the shape of the broad line. First, with the exception of the Fe K$\alpha$ and K$\beta$ narrow features, the line profile is smooth, not suggestive of being composed of a superposition of narrow features. As an example, we replaced the \texttt{relxill} component in our model with two photoionized emission regions (through the \texttt{photemis} model) with a velocity of equal magnitude but opposite sign (i.e. redshift and blueshift), broadened by BH gravity through the \texttt{rdblur} component. We linked all the properties of the two \texttt{photemis} components, and kept the original treatment of the narrow features through \texttt{MYTorus}. This gives a complete model of \texttt{TBabs*[rdblur*(photemis\_blue + photemis\_red) + rdblur*MYTorus]}. This model attempts to explain the shape of the broad line as relativistically broadened emission from a photoionized wind-like structure orbiting the BH. The best-fit using this model produces a fit worse by $\Delta C=87$ for 2 additional degrees of freedom, corresponding to a p-value of $10^{-18}$ obtained through an F-test. Such an explanation to the data is strongly disfavored.

Second, the dashed lines in the two panels in Figure \ref{fig:line_shape} show the line profiles obtained by our model when maintaining the inclination at the best-fit value and varying the spin (left), and when keeping the BH spin constant and varying the inclination (right). We adjusted the model to match the line flux around 7 keV. Note that these curves are not the result of fitting the models to the data. When maintaining the inclination at $\theta=62^\circ$ and varying the BH spin, we plotted the model shape for a non-spinning BH (cyan), intermediate values of 0.5 (magenta) and 0.9 (green), and a nearly maximal spin of 0.99 (orange). Comparing these curves with our best model that measures $a=0.975$ illustrates that lower spins under-predict the shape of the red wing of the line, while higher spins (orange line) overestimate the line width and underestimate its flux. Similarly, when holding the spin constant (right), we illustrate a low $5^\circ$ inclination (cyan), a $27^\circ$ binary inclination measurement of \cite{2011ApJ...742...84O} (magenta), the $\sim45^\circ$ reflection measurement of \cite{2024ApJ...969...40D} (green) and a high $75^\circ$ inclination (orange). Lower inclinations slightly influence the shape of the red wing of the Fe line, but they dramatically under-predict the blue wing. High inclinations broaden the line profile. It is, however, important to note that the simplistic experiment of maintaining one parameter fixed and varying the other does not capture the full extent of the influence on the shape of the Fe line, especially seeing how both parameters act simultaneously on the line profile and given how the MCMC analysis suggests a moderate correlation between the two parameters (see Figure \ref{fig:corner} in Appendix \ref{sec:corner}). However, given how the MCMC analysis suggests $3\sigma$ contours for the two parameters that allow only $60^\circ\leq\theta\leq68^\circ$ and $0.96\leq a\leq 0.99$, this experiment provides a visual explanation of why the model prefers a high but not maximal BH spin of $a=0.98$, and an inclination of $\theta=63^\circ$. For reference, we performed an experiment in which we fixed the BH spin to the maximum allowed value of 0.998, while maintaining the inner disk radius fixed at the ISCO, and refit the spectra. The fit statistic worsens by $\Delta C=24$ for a single parameter.

\section{Results and Discussion}\label{sec:results}

The analysis of XRISM Resolve spectra of the accreting BH HMXB Cygnus X-1 reveals an unprecedented view of the relativistically broadened emission line from the innermost regions around accreting black holes. Owing to the increase in X-ray obscuration during the second half of the exposure, we split the observation around the central time, and extracted spectra from both halves of the observation. Fitting the spectra with a simple power law model highlights residuals that indicate an unequivocal detection of a relativistically broadened Fe K line, and a narrow emission line at the energy of the neutral Fe K$\alpha$ transition. Applying a model that simultaneously describes the continuum, the relativistic reflection, the narrow neutral Fe K$\alpha$ emission line, and the absorption features owing to the onset of the stellar wind of the companion (during the second half of the observation) to the two spectra produces good spectral fits. 

Of the parameters that characterize the shape of the relativistically broadened Fe line, the BH spin, viewing inclination, emissivity profile of the corona, and the ionization of the disk atmosphere produce results consistent between the two spectra. In particular, we measure a BH spin of $a\simeq0.98\pm0.01$ and an inclination of $\theta\simeq63^\circ\pm3^\circ$. Although the uncertainties reported here are simply statistical, the spin measurement is in good agreement with that of \cite{2024ApJ...969...40D}, obtained by combining relativistic reflection measurements from 34 NuSTAR observations. However, the inclination measurement $\theta\simeq63^\circ\pm3^\circ$ is higher than the $\theta=47^\circ\pm10^\circ$ value reported in \cite{2024ApJ...969...40D}, likely pointing to the influence of a high-resolution characterization of the blue wing of the relativistically broadened Fe line, but also at the importance of high-energy spectral coverage of the Compton hump. At the same time, the ability to measure the other parameters that influence the reflection spectrum (Fe abundance, high-energy cutoff, reflection fraction) and the shape of the underlying continuum (power-law index $\Gamma$, normalization of the \texttt{relxill} component, and galactic column density) is likely reduced due to correlations between parameters and limited energy coverage, leading to significantly different measurements resulting from the two halves of the observation. We explored the influence of including spectral coverage of the Compton hump by simultaneously fitting XRISM Resolve and NuSTAR data and found that while many of the parameters in the \texttt{relxill} component can be differently inferred, the measured inclination and BH spin remain consistent. This reinforces the fact that while coverage of the Compton hump is important for breaking parameter degeneracies, the BH spin and inclination are mainly determined by the shape of the spectrum in the Fe band.

The $\theta\simeq63^\circ\pm3^\circ$ inclination measurement obtained from the analysis of the XRISM Resolve spectrum is significantly at odds with the $\sim27^\circ$ orbital inclination of Cygnus X-1 (\citealt{2011ApJ...742...84O}), obtained from optical photometry and spectroscopy. Early optical studies (see, e.g., \citealt{1978SvAL....4..292B}) suggested that the inclination should be in the range of $\sim40^\circ-55^\circ$, and that higher inclinations would likely cause X-ray eclipses. Previous X-ray reflection studies found evidence of misalignment between the inner accretion disk and the orbital plane. For instance, \cite{2014ApJ...780...78T} measured an inclination between $40^\circ$ and $70^\circ$ using NuSTAR and Suzaku spectra, \cite{2015ApJ...808....9P} used updated models to measure $\theta=45.3^\circ\pm0.3^\circ$, and \cite{2016ApJ...826...87W} found an inclination of $\sim40^\circ$ across multiple soft state observations of Cygnus X-1. Such a discrepancy between the inclination of the inner accretion disk and the orbital inclination could lead to the formation of a warped outer region of the accretion disk, and reprocessing of radiation in such a vertical structure could serve as a possible explanation for the origin of the narrow emission lines detected in the spectra. At the same time, X-ray polarization studies using IXPE observations suggest that the viewing inclination must be higher than $65^\circ$ in a sandwich corona geometry or higher than $45^\circ$ for a truncated disk model (\citealt{2022Sci...378..650K}). However, \cite{2001MNRAS.327.1273S} analyzed long-term VLBA and MERLIN radio observations of Cygnus X-1 and concluded that the morphology of the relativistic radio jet detected in the hard state is consistent with either a bending flow or a variable ejection angle. The latter possibility would be indicative of a precessing jet which, if launched along the spin axis of the BH, could either hint at a precessing inner accretion disk in the assumption of alignment with the equatorial plane of the BH, or at a changing coronal geometry. Both effects would lead to a time dependence of reflection-obtained inclination measurements, possibly also correlated with the 150-day variability in the 15-GHz radio flux of Cygnus X-1 (\citealt{1999MNRAS.302L...1P}).

The shape of the narrow Fe K$\alpha$ emission line is best described by X-ray reprocessing (reflection) at relatively large distances through the \texttt{MYTorus} component. We find that the statistically best fit is obtained by convolving the \texttt{MYTorus} reflection with the \texttt{rdblur} component, that describes the relativistic effects around a BH. Although a detailed analysis of the profile of the narrow line is reserved for a future publication, we find that the line prefers a moderate redshift of $z\sim3\times10^{-4}$, corresponding to a velocity of $\sim90\;\rm km/s$. 

%concluding paragraph
The study of the emission signatures in the XRISM Resolve spectrum of the BH HMXB Cygnus X-1 confirms the presence of a relativistically broadened Fe complex rather than a superposition of features that cannot be distinguished at CCD resolution. This broadened feature is consistent with emission originating at the ISCO of the BH. The BH spin measurement obtained using the relativistic reflection method is consistent with past work, reinforcing the robustness of prior BH spin measurements of the source and of the method itself. Future studies that simultaneously leverage the high spectral resolution of Resolve, the high energy coverage of NuSTAR, and the soft energy coverage of Xtend will enable a characterization of the effects of narrow spectral features unaccounted for in prior BH spin measurements using the relativistic reflection method, and unlock a new avenue for exploring the magnitude of systematic uncertainties associated with these measurements. Additionally, future studies of the profile of the narrow Fe K$\alpha$ line will enable a characterization of the geometry of the accretion flow at large radii, and of the mechanism that feeds the accretion disk in this system. Performing similar studies in other accreting BH XBs will unlock a novel and expanding understanding of BH binaries. 

\begin{acknowledgments}
The authors thank the anonymous reviewer, whose comments have significantly improved the paper.
\end{acknowledgments}

\bibliography{sample7}{}
\bibliographystyle{aasjournalv7}

\appendix 
\section{Corner plot}\label{sec:corner}

Here, for completeness, we present the corner plot obtained from the MCMC runs in our analysis. 

\begin{figure}[hb!]
    \centering
    \includegraphics[width=0.95\linewidth]{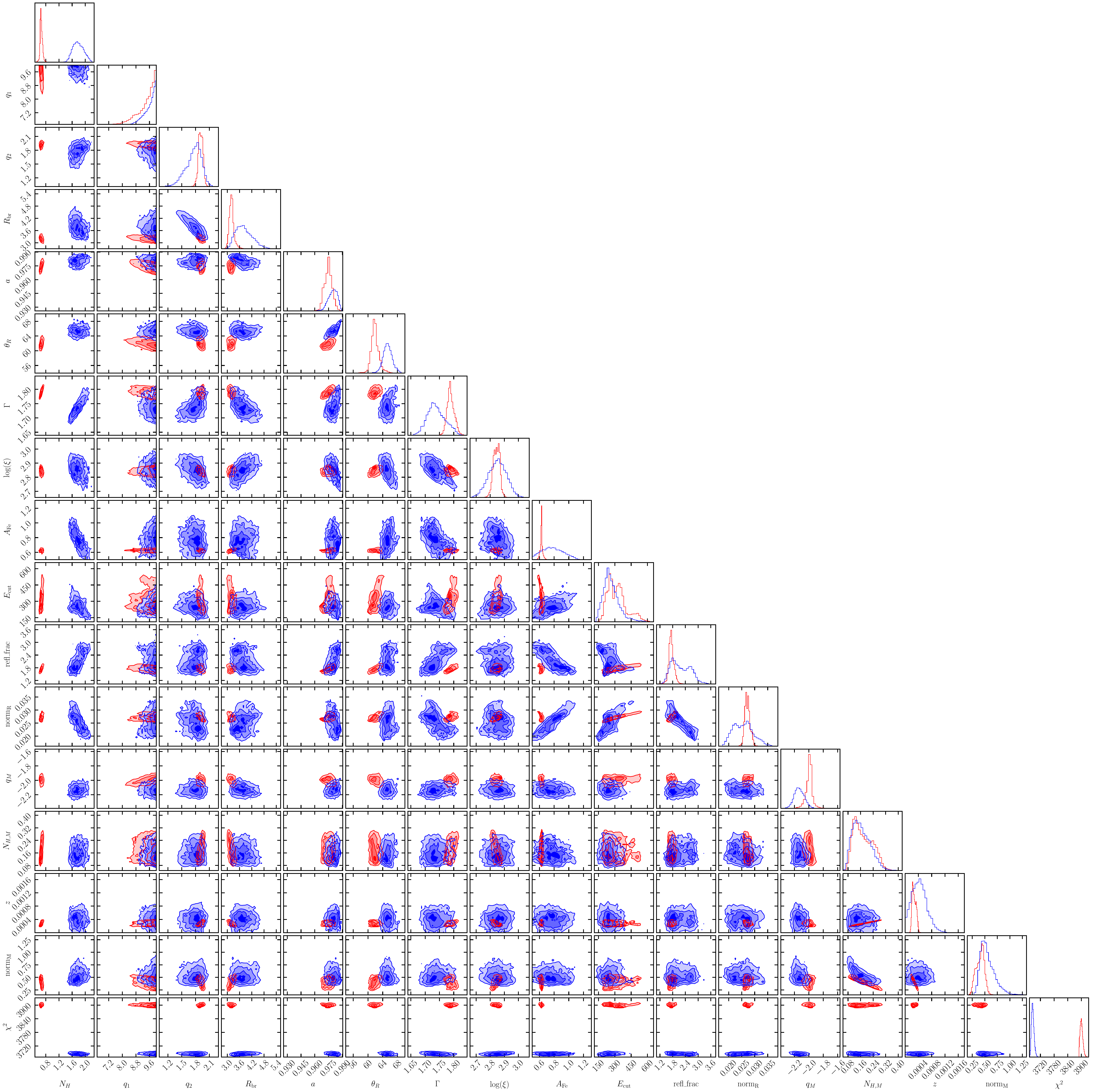}
    \caption{Complete corner plot obtained based on the MCMC runs on the fits to the two halves of the XRISM Resolve observation of Cygnus X-1. The red contours indicate the measurements obtained from the first half the observation, while the blue contours indicate the measurements obtained from the second half.}
    \label{fig:corner}
\end{figure}

%\newpage

\end{document}